\documentclass[conference]{IEEEtran}
\usepackage{cite}

\ifCLASSINFOpdf
 \usepackage[pdftex]{graphicx}
\else
 \usepackage[dvips]{graphicx}
\fi
\usepackage[cmex10]{amsmath}
\usepackage{algorithmic}

\usepackage{array}
\newtheorem{theorem}{Theorem}
\begin{document}
%
\title{Distributed Approximate Message Passing for Compressed Sensing}

\author{\IEEEauthorblockN{Puxiao Han, Ruixin Niu, and Mengqi Ren}
\IEEEauthorblockA{Department of Electrical and Computer Engineering\\
Virginia Commonwealth University, Richmond, VA, 23284, U.S.A.\\
Email: \{hanp, rniu, renm\}@vcu.edu}
}
\maketitle

\begin{abstract}
In this paper, an efficient distributed approach for implementing the approximate message passing (AMP) algorithm, named distributed AMP (DAMP), is developed for compressed sensing (CS) recovery in sensor networks with the sparsity $K$ unknown. In the proposed DAMP, distributed sensors do not have to use or know the entire global sensing matrix, and the burden of computation and storage for each sensor is reduced. To reduce communications among the sensors, a new data query algorithm, called global computation for AMP (GCAMP), is proposed.  The proposed GCAMP based DAMP approach has exactly the same recovery solution as the centralized AMP algorithm, which is proved theoretically in the paper. The performance of the DAMP approach is evaluated in terms of the communication cost saved by using GCAMP. For comparison purpose, thresholding algorithm (TA), a well known distributed Top-K algorithm, is modified so that it also leads to the same recovery solution as the centralized AMP. Numerical results demonstrate that the  GCAMP based DAMP outperforms the Modified TA based DAMP, and reduces the communication cost significantly. 
\end{abstract}
\section{Introduction}
Compressed sensing (CS) has wide applications in various areas of signal processing \cite{CsApp}. Due to the curse of dimensionality, it can be highly demanding to perform CS on a single processor. Further, distributed processing has the potential to reduce communications among distributed sensors.  Hence, distributed CS (DCS) in sensor networks has become an interesting topic. A general DCS system contains two parts: (1) the local computation performed at each sensor, and (2) the global computation to obtain the estimate of the original sparse signal after sensors exchange the results of local computation. Several distributed approaches based on various CS recovery algorithms were proposed. In \cite{DiSP}, a distributed subspace pursuit (DiSP) algorithm  was developed to recover joint sparse signals. In DiSP, each sensor needs to store the global sensing matrix, and  local computation at each sensor  involves  optimization and matrix inversion. The computation and memory burden may become very challenging for each sensor in large-scale problems. Further, in DiSP the sparsity $K$ is assumed to be known, which may not be the case in many applications. In \cite{DBP}, an algorithm named D-ADMM based on basis pursuit (BP) was proposed, in which sensors  do not have to store the entire global sensing matrix. However, each sensor still needs to solve an optimization problem to get an recovery per iteration, and broadcasts it to its neighbors, which may induce high communication cost since the recovery in first few iterations is not sparse. Focusing on these problems, a DCS algorithm based on iterative hard thresholding (IHT) named D-IHT was proposed in \cite{DIHT}. In the local computation, each sensor just performs very simple operations such as matrix transpose, addition and multiplication. In the global computation, thresholding algorithm (TA) \cite{TA} has been applied, which is a popular method to solve the distributed Top-K problem in the field of database querying, to reduce the amount of messages sent between sensors. Nevertheless, in the D-IHT, the sparsity $K$ was also assumed to be known. Further, the D-IHT requires  each local sensor to know certain prior knowledge about the global sensing matrix, such as its $L_2$ norm. For a certain sensor node (or a fusion center) to know the global sensing matrix to calculate and then broadcast its $L_2$ norm, each of the rest sensor nodes has to either transmit its local sensing matrix or the seed of its local random number generator used to generate the corresponding local sensing matrix.   

In this paper, we do not assume the knowledge of sparsity and hence the IHT cannot be directly applied. Instead, we propose a distributed algorithm based on approximate message passing (AMP) \cite{AMP}, which does not require any prior knowledge of the sparse signal, and has a linear convergence rate \cite{AMP}, \cite{convergence}. For the proposed  distributed AMP (DAMP) approach, we do not assume any prior knowledge of the global sensing matrix. Distributed sensors do not need to store the entire global sensing matrix. In the local computation, each sensor only performs simple matrix operations, and in the global computation per iteration, we propose a new algorithm, Global Computation for AMP (GCAMP), to reduce the amount of data transmitted in the sensor network. To the best of our knowledge, the proposed approach is the first distributed AMP algorithm ever developed.


\section{DAMP System}
\subsection{The Original AMP}
A task of CS is to recover a sparse signal $s_0 \in R^N$ from its measurement $y = As_0 + n$, where $A \in R^{M \times N}$ is the sensing matrix and $n$ is an additive noise, by solving the problem:
\begin{equation}
\label{eqn5}
\min_{x} \frac{1}{2}||y-Ax||^2_2 + \lambda ||x||_1
\end{equation}
where $\lambda>0$ is the regularization parameter. However, $\lambda$ is not given in most practical cases. AMP is a good solution to the problem  \cite{AMP} without prior knowledge about $K$ and $\lambda$. Starting from $x_0 = 0$ and $z_0 = y$, it recursively gets the new estimate of $s_0$ as follows:
\begin{equation}
\label{eqn6}
x_{t+1}= \eta_{t}(x_t+A^{T}z_t;\tau\sigma_t)
\end{equation}
\begin{equation}
\label{eqn7}
z_{t+1}= y - Ax_{t+1} + \frac{||x_{t+1}||_0}{M}z_{t}
\end{equation}
where $\left[\cdot\right]^T$ denotes the transpose operation, $||\cdot||_0$ is the $l_0$ norm of a vector, ${\sigma}^{2}_t = \frac{||z_t||^2}{M} $ \cite{TUNE}, 
\begin{equation}
\label{eqn9}
\eta_{t}(x;\beta)=
\begin{cases}
(|x| - \beta)\text{sgn}(x), & |x| > \beta \\
0, & |x| \leq \beta
\end{cases}
\end{equation}
 and $\tau$ is a parameter whose optimal value depends on $\kappa=\frac{M}{N}$ and $\rho=\frac{K}{M}$ \cite{TUNE}. Since $K$ is unknown, a tuning procedure is needed, which will be presented later in this paper, to find a value for $\tau$ which is very close to the optimum.


\subsection{The Distributed Framework of AMP}
Let us consider a sensor network with $P$ distributed sensors. Each sensor $p$ ($p=1$, $\cdots$, $P$) takes a measurement of $s_0$ as
\begin{equation}
\label{eqn10}
\left[
\begin{matrix}
y^1 \\
\vdots \\
y^P
\end{matrix}
\right] = \left[
\begin{matrix}
A^1 \\
\vdots \\
A^P
\end{matrix}
\right]s_0 + 
\left[
\begin{matrix}
n^1 \\
\vdots \\
n^P
\end{matrix}
\right]
\end{equation}
Then,  \eqref{eqn6} and \eqref{eqn7} can be re-written as:
\begin{equation}
\label{eqn11}
x_{t+1}= \eta_{t}\left(x_t+\Sigma_{p=1}^P A^{pT} z^p_t;\tau{\sigma_t}\right)
\end{equation}
\begin{equation}
\label{eqn12}
z^p_{t+1}=y^p-A^p x_{t+1}+\frac{||x_{t+1}||_0}{M}z^p_t, \forall p = 1,\cdots,P
\end{equation}
 By introducing an intermediate matrix $W_t=\left[w^1_t, \dots, w^P_t\right]$ with each column computed by the corresponding sensor as:
\begin{equation}
\label{eqn13}
w^p_t=
\begin{cases}
x_t + A^{pT}z^p_t, & p=1 \\
A^{pT}z^p_t, & \mbox{otherwise}
\end{cases}
\end{equation}
which is similar to that in \cite{DIHT}, \eqref{eqn11} becomes
\begin{equation}
\label{eqn14}
x_{t+1}=\eta_t\left(\Sigma_{p=1}^{P}w^p_t;\tau{\sigma}_t\right)
\end{equation}
Therefore, DAMP can be divided into two parts: local computation of $z^p_t$ and $w^p_t (p=1$, $\cdots$, $ P)$,  and global computation of $x_{t+1}$ and $\sigma_{t+1}$, in which transmission of data between sensors is needed. For the latter, a natural approach is to send all the data in $w^p_t (p=2,\cdots, P)$ to sensor 1, which induces a high communication cost when $N$ is large. Therefore, how to reduce the communication cost, meanwhile maintaining the same recovery solution as the  centralized AMP, is the main focus of this paper.

\subsection{GCAMP Algorithm}
Let us  denote $v(n)$ as the $n$-th component of a vector $v$. According to \eqref{eqn14}, $x_{t+1}(n)=0$ if $|\Sigma_{p=1}^P w^p_t(n)| \leq \beta = \tau\sigma_t$. Therefore, we only need to know all the $n$s such that $|\Sigma_{p=1}^P w^p_t(n)| > \beta$ in the global computation. This is similar to Top-K problem in the field of distributed database querying, which is to find the $K$ largest components of $\Sigma_{p=1}^P w^p_t$. In \cite{TPUT} the three-phase uniform threshold (TPUT) algorithm, an efficient approach to solve the Top-K problem with a known $K$, is proposed. However,  our problem is different from  the Top-K problem. First, we do not know how many components of $\Sigma_{p=1}^{P}w^{p}_{t}$ have magnitude larger than $\beta$; second, TPUT requires  $w^p_t(n)$'s  to be non-negative, while in our problem, they can be any real numbers. Hence, TPUT cannot be applied in our case. Nevertheless, it does provide some insight on how to design the communication algorithm in distributed systems. Here, we propose the GCAMP algorithm which is shown in Table \ref{tb:GCAMP}. 
\begin{table}[hbt]
\caption{ GCAMP Algorithm }
\begin{center}
\label{tb:GCAMP}
    \begin{tabular}{  l }
   \textbf{Input} $w^1_t,\cdots , w^P_t$, $\beta = \tau \sigma_{t}$; \\ \\ \hline 
\\ \textbf{Step I} Set $T=\beta\theta/(P-1)$, where $\theta \in (0, 1)$ is a tuned parameter;
\\for sensor $p$ = 2:$P$\\
\hspace{3mm} denote $R_p = \{n:|w^p_t(n)|>T\}$; \\
\hspace{3mm} send all ($n, w^p_t(n)$) pairs for $n \in R_p$ to sensor 1; \\
endfor \\
\textbf{Step II} for sensor 1, define $I_S(x) := 1$ if $x \in S$; 0 o.w;  \\
for $n$ = 1:$N$ \\
\hspace{3mm} get $S_n := \{p=2,\cdots,P:I_{R_p}(n)=1\}$ with cardinality $m_n$; \\
\hspace{3mm} Compute $U(n)=|w_t^{1}(n)+\Sigma_{p \in S_n}(w^p_t(n))|+(P-1-m_n)T$; \\
\hspace{3mm} if $U(n)>\beta$ and $m_n<P-1$ \\
\hspace{6mm} broadcast the index $n$ to other sensors; \\
\hspace{3mm} endif \\
endfor \\
\textbf{Step III} denote $F = \{n: U(n)>\beta$, $m_n<P-1\}$;\\
for sensor $p$ = 2:$P$ \\
\hspace{3mm} send all ($n, w^p_t(n)$) pairs for $n \in F\backslash R_p$ to sensor 1; \\
endfor \\ 
\textbf{Step IV} for sensor 1, initialize $x_{t+1}=0$; \\
for $n \in V:=\{n:U(n)>\beta\}$ \\
\hspace{3mm} Update $x_{t+1}(n) = \eta_t\left(\Sigma_{p=1}^{P}w^p_t(n);\beta\right)$ by \eqref{eqn9}; \\
endfor \\ \\ \hline
\\ \textbf{Output} $x_{t+1}$ \\ \\
 \hline
     \end{tabular}
\end{center}
\end{table}
\begin{theorem}
\label{thm:GCAMP}
In each iteration, $U(n)$ is an upper bound of $\left|\Sigma_{p=1}^P w^p_t(n)\right|$ for all $n$, and the $x_{t+1}$ which GCAMP algorithm obtains (denoted as $x^G_{t+1}$) is exactly the same as that obtained by the original centralized AMP algorithm (denoted as $x^A_{t+1}$).
\end{theorem}

\noindent \underline{\em Proof:} For any $n=1,\cdots,N$, we have
\begin{eqnarray}
\Sigma_{p=1}^P w^p_t(n) = w^1_t(n) + \Sigma_{p \in S_n} w^p_t(n) + \Sigma_{p \geq 2, p \notin S_n} w^p_t(n) 
\end{eqnarray}
Then, applying the triangle inequality, we have
\begin{eqnarray}
\label{eqn:upper}
\left|\Sigma_{p=1}^P w^p_t(n)\right| \leq \left|w^1_t(n) + \Sigma_{p \in S_n} w^p_t(n)\right| + \left|\Sigma_{p \geq 2, p \notin S_n} w^p_t(n)\right| \\ \nonumber
 \leq \left|w^1_t(n) + \Sigma_{p \in S_n} w^p_t(n)\right| + (P-1-m_n)T = U(n) 
\end{eqnarray}
$\forall n \notin V$, $x^G_{t+1}(n) = 0$; by \eqref{eqn:upper}, $|\Sigma_{p=1}^P w^p_t(n)| \leq U(n) \leq \beta$, so $x^A_{t+1}(n) = 0$. $\forall n \in V$, $x^G_{t+1}(n)=x^A_{t+1}(n)=\eta_t(\Sigma_{p=1}^{P}w^p_t(n);\beta)$. Therefore, $x^G_{t+1}=x^A_{t+1}$. 

\begin{figure}[!t]
\centering
\includegraphics[width=3.45 in]{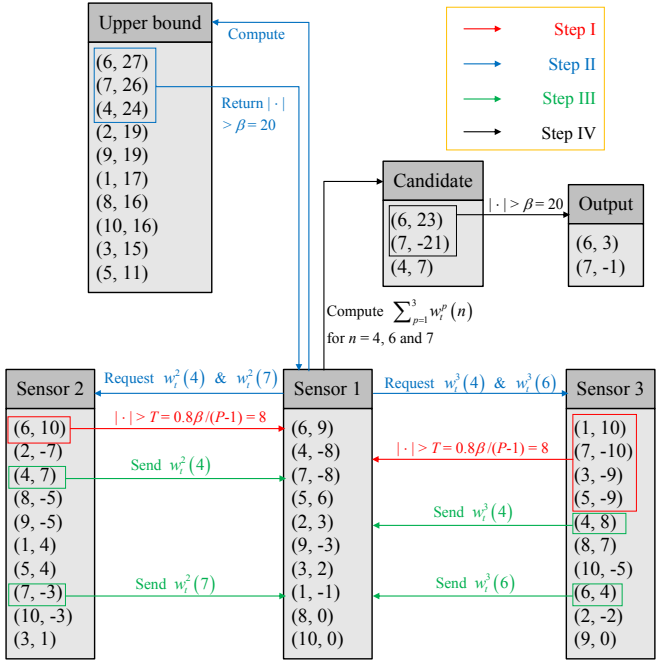}
\caption{An example of GCAMP algorithm}
\label{fig:GCAMP}
\end{figure}

In Fig. \ref{fig:GCAMP}, an example is provided to illustrate how GCAMP works, in which each sensor $p$ already sorts $w^p_t(n)$ in descending order of magnitudes, and stores the data in the form of ($n$, $w^p_t(n)$) pairs ($p=1,\cdots,3,n=1,\cdots,10$). Suppose $\beta = 20$ and $\theta = 0.8$, since we have $P = 3$ sensors, we get $T = \beta\theta/(P-1) = 8$. In step I, sensors 2 to $P$ send all ($n$, $w^p_t(n)$) pairs with $|w^p_t(n)| > T$ to sensor 1. In step II, sensor 1 receives the data, computes upper bounds $U(n)$ for $n = 1,\cdots,10$ and obtains $F = V = \{4,6,7\}$. Then sensor 1 broadcasts indices in $n \in F$. In step III, sensor 2 sends $w^2_t(4)$ and $w^2_t(7)$, and sensor 3 sends $w^3_t(4)$ and $w^3_t(6)$ to sensor 1. Finally, in step IV, sensor 1 computes $x_{t+1}(n)$ for $n \in V$ by \eqref{eqn13}, and outputs the non-zero components of $x_{t+1}$. Overall, in this example, only 9 data points are sent from other sensors to sensor 1, and the total number of messages is 12 (9 data points plus 3 broadcast requests).
\subsection{Tuning of $\tau$ Values}
With the GCAMP algorithm, DAMP can be developed. We adopt the tuning framework in \cite{CAMPDetect} to find the optimal value for $\tau$. First, a descending candidate list of candidate values of $\tau$, $\{\tau\}_{l=1}^{L} := [\tau_\text{max},\cdots, \tau_\text{max}-(l-1)\Delta \tau,\cdots,\tau_\text{max}-(L-1)\Delta \tau]$ is generated. Then, for each candidate $\tau_l$, we run iterations in \eqref{eqn11} and \eqref{eqn12} until $x_{t}$ and $\sigma_t$ converge to $x_l^{*}$ and $\sigma_l^*$, and use them as the initial estimates for the iterations using the next candidate $\tau_{l+1}$. We repeat this process until $\sigma_l^*$ is not decreasing, and get the optimal $\tau$ value as well as the final estimate of $s_0$. The pseudo code of DAMP algorithm is shown in Table \ref{tb:DAMP}.\\
How to choose the maximum candidate value, i.e., $\tau_\text{max}$, is also an interesting problem. In \cite{CAMPDetect}, the authors set $\tau_\text{max}=\frac{||A^T y||_{\infty}}{\sqrt{||y||^2_2/M}}$, which $||\cdot||_\infty$ is the magnitude of the largest-in-magnitude component in a vector. Denote $\tilde{x}_{t}:=x_t + A^T z_t=\Sigma_{p=1}^{P}w^{p}_{t}$, since at the beginning, $x_0=0$ and $z_0=y$, we have $\tilde{x}_{0}=A^T y$ and $\sigma_0=\sqrt{\frac{||y||^2_2}{M}}$. Therefore, $\forall n=1,\cdots,N$, we have $|\tilde{x}_{0}(n)|\leq \tau_\text{max}\sigma_0$. This implies that the optimal value for $\tau$ cannot be greater than $\tau_\text{max}$. Here, we propose a different approach. According to \cite{Dynamic}, as $N \rightarrow \infty$, asymptotically each component of $\tilde{x}_{t}-s_0$  is independent and identically distributed (i.i.d.) random variable, following a ${\cal N}(0,\sigma^2_t)$ distribution. Therefore, we can build a ($1 - \alpha$) confidence interval (CI) $\left[-z_{\frac{\alpha}{2}}\sigma_t, z_{\frac{\alpha}{2}}\sigma_t\right]$, where $z_{\alpha}$ is defined such that $\frac{1}{\sqrt{2\pi}}\int_{z_{\alpha}}^{+\infty}\exp(-\frac{t^2}{2})dt=\alpha$. Hence, $\forall n = 1,\cdots,N$, if $s_0(n)=0$, with probability $1-\alpha$, $\tilde{x}_{t}(n)$ will be in the CI; on the other hand, if for some $n$, $|\tilde{x}_{t}(n)|>z_{\frac{\alpha}{2}}\sigma_t$, then with probability at least $1-\alpha$, $s_0(n)$ is a non-zero component. Therefore, we can choose a very small $\alpha$, and let $\tau_\text{max}=z_{\frac{\alpha}{2}}$. For example, we can let $\alpha=0.0027$ and $\tau_\text{max}=z_{\frac{\alpha}{2}}=3$.\\
Note that in every iteration involving \eqref{eqn11} and \eqref{eqn12}, after GCAMP returns $x_{t+1}$, sensor 1 broadcasts non-zero components of $x_{t+1}$ as well as their indices. In DAMP, we tune the optimal $\tau$ value in a descending order, which implies a larger threshold $\beta=\tau\sigma_t $ in the beginning. Therefore, different from \cite{DBP}, we have a sparse estimate $x_{t+1}$ even at the first few iterations. Hence, the communication cost for broadcasting $x_{t+1}$ is negligible compared with that of GCAMP. Once knowing $x_{t+1}$, each local sensor can obtain $z^p_{t+1}$ using \eqref{eqn12} and $\sigma^p_{t+1} = ||z^p_{t+1}||_2$ ($ p=1$, $\cdots$, $ P$). Next, each sensor $p \geq 2$ just sends a scalar $\sigma^p_{t+1}$ to sensor 1, which needs $P-1$ messages. Then, sensor 1 computes ${\sigma}_{t+1} = \sqrt{\Sigma_{p=1}^{P}(\sigma^p_{t+1})^2/M}$, updates $\beta$ and $T$, and broadcasts the scalar $T$ to other sensors. Overall, GCAMP incurs most of the communication cost in DAMP.
\begin{table}[hbt]
\caption{ DAMP Algorithm }
\begin{center}
\label{tb:DAMP}
    \begin{tabular}{  l }
   \textbf{Input} $\{y\}_{p=1}^{P}$, $\{A\}_{p=1}^{P}$, $\{\tau\}_{i=1}^{L}$, maxiter, $\epsilon$; \\ \\ \hline \\
\textbf{Initialization} $x_0 = 0, z^p_0 = y^p \text{ for } p=1\cdots P, \sigma_0 = \sqrt{\Sigma_{p=1}^{P}||z^p_0||^2_2/M}$; \\
for $i$ = 1:$L$ \\
\hspace{3mm} for $t$ = 1:maxiter \\
\hspace{6mm}	for $p$ = 1:$P$  \\
\hspace{9mm}			Compute $w^p_{t-1}$ by \eqref{eqn13}; \\
\hspace{6mm}		endfor \\
\hspace{6mm}		$x_t = \text{GCAMP}(w^1_{t-1},\cdots ,w^P_{t-1}, \beta=\tau_{i}\sigma_{t-1})$; \\
\hspace{6mm}		for $p$ = 1:$P$  \\
\hspace{9mm}			Compute $z^p_t$ by \eqref{eqn12}; \\
\hspace{6mm}		endfor \\
\hspace{6mm}		$\sigma_t = \sqrt{\Sigma_{p=1}^{P}||z^p_t||^2_2/M}$ \\
\hspace{6mm}		if $|\sigma_t-\sigma_{t-1}| < \epsilon \sigma_{t-1}$ \\
\hspace{9mm}			$\sigma(\tau_i) = \sigma_t, x(\tau_i) = x_t, z^{p}(\tau_i) = z^p_t \text{ for } p=1\cdots P$; \\
\hspace{9mm}			break; \\
\hspace{6mm}		endif \\
\hspace{3mm}	endfor \\
\hspace{3mm}	if $\sigma(\tau_i)>\sigma(\tau_{i-1})$ \\
\hspace{6mm}		$\tau^*={\tau_{i-1}}, \sigma^*=\sigma(\tau^*), x^*=x(\tau^*)$; \\
\hspace{6mm}		return; \\
\hspace{3mm}	else \\
\hspace{6mm}		$\sigma_0 = \sigma(\tau_i), x_0 = x(\tau_i), z^p_0 = z^{p}(\tau_i) \text{ for } p=1\cdots P$; \\
\hspace{3mm}	endif \\
endfor \\ \\ \hline
\\ \textbf{Output} $\tau^*,\sigma^*,x^*$ \\ \\
 \hline
     \end{tabular}
\end{center}
\vspace{-0.14in}
\end{table}
\subsection{Comparison of GCAMP and Modified TA}
TA \cite{TA} is another popular algorithm solving Top-K problems. Similar to TPUT, TA also requires the knowledge of  $K$ and all entries in $W_t$ to be non-negative. Therefore, we propose a modified TA algorithm as in Table \ref{tb:TA}, and let it be a control algorithm for GCAMP.
\begin{table}[hbt]
\caption{ Modified TA Algorithm }
\begin{center}
\label{tb:TA}
    \begin{tabular}{  l }
 \textbf{Input} $w^1_t,\cdots , w^P_t$, $\beta = \tau \sigma_{t}$; \\ \\  \hline
\\ \textbf{Initialization} $x_{t+1}=0, N_s = 0$;  \\
for sensor $p$ = 1:$P$ \\
\hspace{3mm} sort components of $w^p_t$ in descending order of magnitudes; \\
\hspace{3mm} define the sorted vector as $s^p_t$ and $I^p_t(n):=l$ s.t. $w^p_t(l) = s^p_t(n)$; \\
\hspace{3mm} mark all ($I^p_t(n)$, $s^p_t(n)$) pairs as ``unsent'';\\
endfor \\
while 1 \\
\hspace{3mm} for $p$ = 1:$P$, do the following process named global summation\\
\hspace{6mm} find the first ($I^p_t(n)$, $s^p_t(n)$) pair marked ``unsent'' from top; \\
\hspace{6mm} set $u_p = s^p_t(n)$, broadcast ($I^p_t(n)$, $u_p$) to other sensors; \\
\hspace{6mm} mark ($I^p_t(n)$, $s^p_t(n)$) as ``sent''; \\
\hspace{6mm} for sensor $q \neq p$ \\
\hspace{9mm} store $u_p$ and send ($I^p_t(n)$, $w^q_t(I^p_t(n))$) to sensor $p$;  \\
\hspace{9mm} mark ($I^p_t(n)$, $w^q_t(I^p_t(n))$) as ``sent''; \\
\hspace{6mm} endfor \\
\hspace{6mm} update $x_{t+1}(I^p_t(n))=\eta_t(\Sigma_{p=1}^P w^p_t(I^p_t(n));\beta)$; \\
\hspace{6mm} number of global summations $N_s = N_s + 1$;\\
\hspace{6mm} if $N_s \geq P$ and $\Sigma_{p=1}^P |u_p| \leq \beta$, or if $N_s \geq N$ \\
\hspace{9mm} 	the algorithm terminates; \\
\hspace{6mm} endif \\
\hspace{3mm}endfor \\
endwhile \\ \\
 \hline
 \\ \textbf{Output} $x_{t+1}$ \\ \\
 \hline
     \end{tabular}
\end{center}
\end{table}
\begin{theorem}
\label{thm:TA}
In each iteration, Modified TA algorithm also gives exactly the same $x_{t+1}$ as that of original AMP algorithm.
\end{theorem}
\noindent \underline{\em Proof:} Modified TA is composed of a series of global summation, where a global summation means computing $|\Sigma_{p=1}^P w^p_t(n)|$ for some $n$. $N_s$ is a counter recording the number of global summations. At the very end of one global summation, for each $n$, either the $(n, w^p_t(n))$ pairs for all $p$ are marked as ``sent"; or none of them are marked as ``sent". So we can just say $n$ is marked as ``sent" or not. It is easy to show that, $\Sigma_{p=1}^P |u_p|$ is an upper bound of $|\Sigma_{p=1}^P w^p_t(n)|$ for all $n$ that have not been marked as ``sent"; if $\Sigma_{p=1}^P |u_p| \leq \beta$, then we have $|\Sigma_{p=1}^P w^p_t(n)| \leq \beta$ for these $n$. As the algorithm terminates, we do not lose any non-zero components of $x_{t+1}$.

\noindent \underline{\em Number of Messages:}
For a set, denote $|\cdot|$ as its cardinality. For GCAMP, the total number of messages is $\Sigma_{p=1}^{P} |R_p|+|F|+\Sigma_{p=1}^{P}|F\backslash R_p|$; for Modified TA, in each global summation, there are 1 broadcasting message from some sensor to others and $P-1$ incoming messages, so the total number of messages is $PN_s$. It is easy to check that, for the data set in Figure $\ref{fig:GCAMP}$, Modified TA needs $PN_s = 3\times 9 = 27$ messages, more than twice of that of GCAMP.
\section{Numerical Results}
\subsection{Performance Measures}
Since we have proved that the DAMP algorithm has exactly the same solution as the original AMP, and the recovery accuracy and convergence of AMP has been well studied in literature, it is not necessary to evaluate them  again in the paper. Instead, as DAMP is a distributed algorithm, it is important to evaluate the communication cost saved by using GCAMP. So we use the number of messages transmitted as the performance measure, which is widely used in literature \cite{TA,TPUT}. We compare the number of messages used in GCAMP to that in Modified TA. Considering the approach sending all data to sensor 1, which has a total number of messages $N(P-1)$, we define normalized message number  (NMN) as
\begin{equation}
\mu_{M} = \frac{\text{number of messages in computing }x_{t+1}}{N(P-1)}
\end{equation}
which is $\mu_{M}=\frac{\Sigma_{p=1}^{P} |R_p|+|F|+\Sigma_{p=1}^{P}|F\backslash R_p|}{N(P-1)}$ for GCAMP and $\mu_{M}=\frac{N_s P}{N(P-1)}$ for Modified TA.
\subsection{Simulation Setup}
Our focus is  not to investigate large-scale problems, but to develop distributed algorithms and evaluate their efficiency in reducing communication costs. Nevertheless, we still use a considerably large $N = 5000$, and choose $\kappa $ from $[0.1,\;0.5]$, $\rho$ from $[0.1,\;0.3]$, which leads to $M=N\kappa $ in $[500,\;2500]$ and $K=M\rho $ in [50, 750]. The problem scales used in our paper is larger than those used in other DCS publications  \cite{DIHT}. The number of sensors $P$ is within $[5,\;50]$. The sensing matrix $A$ with i.i.d. entries  $\sim$ ${\cal N}(0,\frac{1}{M})$ is partitioned into $P$ parts with each sensor having a $(M/P) \times N$ submatrix. Each component of $s_0$ is i.i.d. drawn from
\begin{equation}
f_X(x) = \kappa\rho G(x) + (1-\kappa\rho) \delta(x)
\end{equation}
where $G(x)$ is the probability density function (pdf) of the standard Gaussian distribution and $\delta(x)$ is the Dirac Delta function. The measurements of $s_0$ are corrupted by an additive noise $n\sim {\cal N}(0,\sigma^{2}I_M)$ and $\sigma$ is the standard deviation with a value in [0.01, 0.1]. The parameter $\theta$ in GCAMP is set to 0.8. Regarding the tuning procedure for optimal $\tau$ values, we make a candidate list for $\tau$ of length 11, starting from 3 with a step -0.2; for each candidate, the convergence criteria is $|\sigma_t-\sigma_{t-1}| < 0.01 \sigma_{t-1}$.  We compare $\bar{\mu}_M$ defined as $\mu_M$ averaged over iterations based on $100$ Monte-Carlo runs. 
\subsection{Performance Evaluation}
\begin{figure}[!t]
\centering
\includegraphics[width=3.45 in]{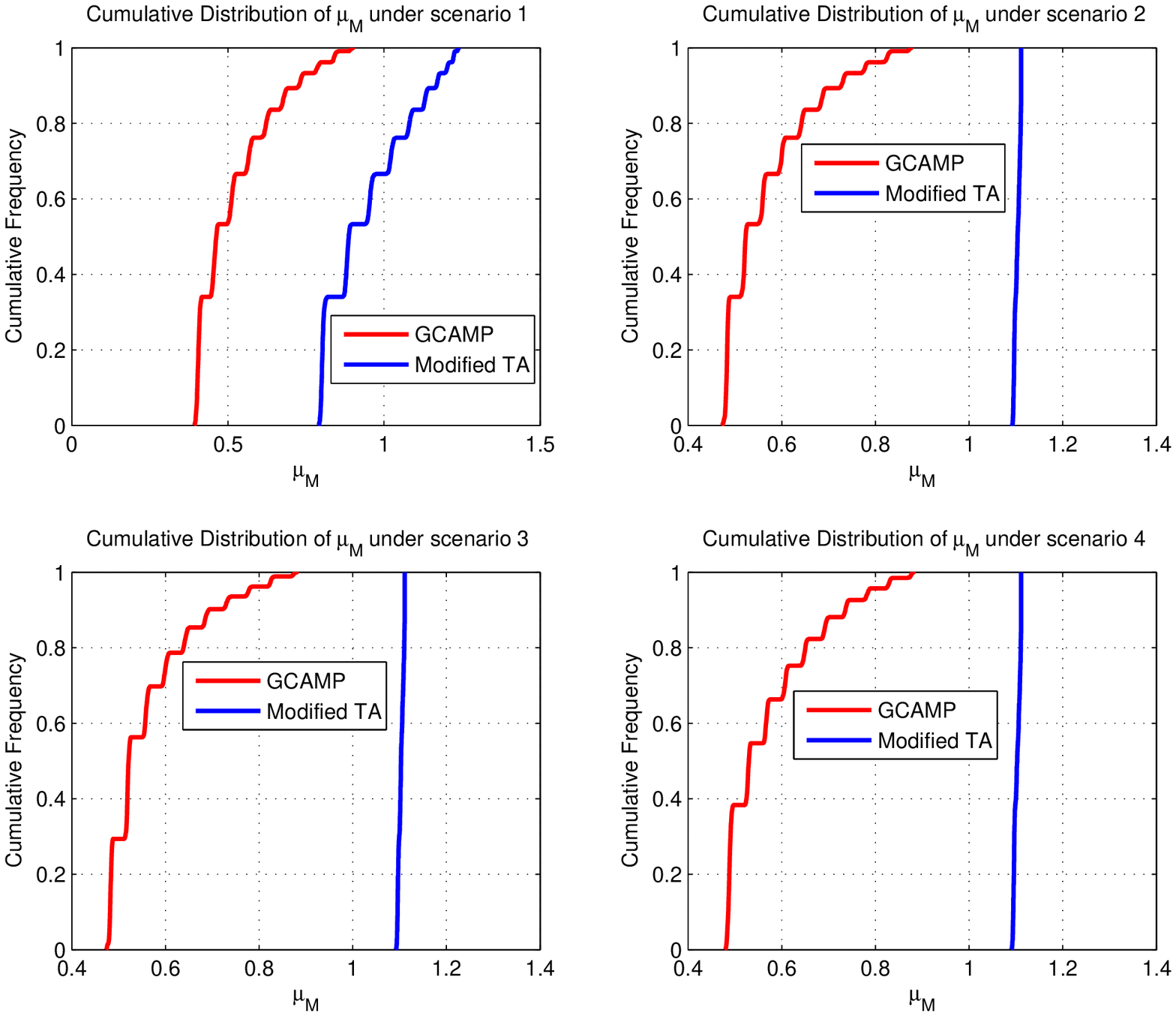}
\caption{Cumulative distributions of $\mu_M$ for GCAMP and Modified TA}
\vspace{-0.25in}
\label{fig:cdf}
\end{figure}

We evaluate $\bar{\mu}_M$ in three settings:
I) fix $\sigma=0.02$ and $P=10$, and change the values of $\kappa$ and $\rho$;
II) fix $\kappa=0.2$, $\rho=0.1$ and $P=10$, and change the values of $\sigma$; 
III) fix $\kappa=0.2$, $\rho=0.1$ and $\sigma=0.02$, and change the values of $P$.
Tables \ref{tb:kappa_rho}, \ref{tb:sigma} and \ref{tb:P} show the corresponding numerical results for I), II) and III) respectively. In the tables, the former entry  in each pair inside the parentheses denotes $\bar{\mu}_M$ for GCAMP, and the latter denotes that for Modified TA. It is clear that in each case, GCAMP outperforms Modified TA significantly. Modified TA always uses more messages than $N(P-1)$ except for the case $P=5$, while GCAMP can save the number of messages from 22.7\% to 48.2\%.
Fig. \ref{fig:cdf} gives the cumulative distributions of $\mu_M$ in each iteration for GCAMP and Modified TA under 4 different scenarios: 1) $\kappa=0.2, \rho=0.1, \sigma=0.02, P=5; $ 2) $\kappa=0.2, \rho=0.1, \sigma=0.02, P=10; $ 3) $\kappa=0.2, \rho=0.1, \sigma=0.01, P=10; $ 4) $\kappa=0.3, \rho=0.1, \sigma=0.02, P=10 $. It provides us much more detailed information on the distribution of  $\mu_M$ for each algorithm. It is clear that under each scenario, Modified TA uses more than $N(P-1)$ messages in at least 33.4 $\%$ of the total iterations; while GCAMP never uses more than $0.91N(P-1)$ messages in any iteration, and among more than $95 \%$ of the total iterations, it just uses $\left[40\%,80\%\right]\times N(P-1)$ messages, that is, it can save $20\%\sim 60\%$ of the messages with probability at least $95\%$.
\begin{table}[hbt]
\caption{ $\bar{\mu}_M$ for GCAMP and Modified TA with  Different $\kappa$ and $\rho$ }
\begin{center}
\label{tb:kappa_rho}
    \begin{tabular}{ | r | p{1cm} | p{1cm} | p{1cm} | p{1cm} | p{1cm} |}
    \hline
 &  $\kappa=0.1$ & 0.2  & 0.3 & 0.4 & 0.5 \\ \hline
$\rho$=0.10&(0.547, 1.101)&(0.567, 1.103)&(0.573, 1.103)&(0.587, 1.103)&(0.589, 1.103) 
 \\ \hline
0.15 &(0.621, 1.108)&(0.616, 1.106)&(0.632, 1.107)&(0.635, 1.107)&(0.639, 1.106) 
 \\ \hline
0.20 &(0.659, 1.108)&(0.667, 1.108)&(0.672, 1.108)&(0.691, 1.109)&(0.684, 1.108) 
 \\ \hline
0.25 &(0.651, 1.107)&(0.689, 1.109)&(0.707, 1.109)&(0.725, 1.109)&(0.731, 1.109) 
 \\ \hline
0.30 &(0.632, 1.108)&(0.690, 1.109)&(0.737, 1.109)&(0.751, 1.110)&(0.755, 1.110)
 \\
 \hline
     \end{tabular}
\end{center}
\vspace{-0.2in}
\end{table}

\begin{table}[hbt]
\caption{ $\bar{\mu}_M$ for GCAMP and Modified TA with Different $\sigma$ }
\begin{center}
\label{tb:sigma}
    \begin{tabular}{ | p{1.2cm} | p{1.2cm} | p{1.2cm} | p{1.2cm} | p{1.2cm} |}
    \hline
  $\sigma=0.01$ & 0.02  & 0.03 & 0.04 & 0.05 \\ \hline
(0.564, 1.103) &	(0.567, 1.103) &	(0.574, 1.104) &	(0.576, 1.104) &	(0.582, 1.104)
 \\ \hline
$\sigma=0.06$ & 0.07  & 0.08 & 0.09 & 0.1 \\ \hline
(0.583, 1.104) &	(0.589, 1.104) &	(0.590, 1.104) &	(0.592, 1.105) &	(0.590, 1.105)
 \\ 

 \hline
     \end{tabular}
\end{center}
\vspace{-0.2in}
\end{table}

\begin{table}[hbt]
\caption{ $\bar{\mu}_M$  for GCAMP and Modified TA with Different $P$ }
\begin{center}
\label{tb:P}
    \begin{tabular}{ | p{1.2cm} | p{1.2cm} | p{1.2cm} | p{1.2cm} | p{1.2cm} |}
    \hline
  $P=5$ & 10  & 15 & 20 & 25 \\ \hline
(0.518, 0.941) &	(0.567, 1.103) &	(0.623, 1.071) &	(0.664, 1.053) &	(0.694, 1.042)
 \\ \hline
$P=30$ & 35  & 40 & 45 & 50 \\ \hline
(0.717, 1.034) &	(0.735, 1.029) &	(0.751, 1.026) &	(0.763, 1.023) &	(0.773, 1.020)
 \\ 

 \hline
     \end{tabular}
\end{center}
\vspace{-0.2in}
\end{table}

\section{Conclusion}
Assuming the sparsity of the original signal to be unknown,  the DAMP approach has been developed for performing compressed sensing in distributed sensor networks, consisting a series of local and global computations. We proposed the GCAMP in the stage of global computation to reduce the number of messages per iteration, and proved theoretically that DAMP based on GCAMP has exactly the same solution as the original AMP. Meanwhile, we modified TA algorithm so that it can be used in DAMP, which also has exactly the same solution as the original AMP, and used it as the control algorithm for GCAMP in evaluating the communication cost savings. Numerical results demonstrated that GCAMP based DAMP outperforms Modified TA based DAMP significantly, and is very efficient in reducing communication costs.

\bibliographystyle{IEEEtran}
\bibliography{DAMPREF}

\begin{thebibliography}{10}
\providecommand{\url}[1]{#1}
\csname url@samestyle\endcsname
\providecommand{\newblock}{\relax}
\providecommand{\bibinfo}[2]{#2}
\providecommand{\BIBentrySTDinterwordspacing}{\spaceskip=0pt\relax}
\providecommand{\BIBentryALTinterwordstretchfactor}{4}
\providecommand{\BIBentryALTinterwordspacing}{\spaceskip=\fontdimen2\font plus
\BIBentryALTinterwordstretchfactor\fontdimen3\font minus
  \fontdimen4\font\relax}
\providecommand{\BIBforeignlanguage}[2]{{%
\expandafter\ifx\csname l@#1\endcsname\relax
\typeout{** WARNING: IEEEtran.bst: No hyphenation pattern has been}%
\typeout{** loaded for the language `#1'. Using the pattern for}%
\typeout{** the default language instead.}%
\else
\language=\csname l@#1\endcsname
\fi
#2}}
\providecommand{\BIBdecl}{\relax}
\BIBdecl

\bibitem{CsApp}
M.~Duarte and Y.~Eldar, ``{Structured compressed sensing: From theory to
  applications},'' \emph{IEEE Trans. Sig. Proc.}, vol.~59, pp. 4053--4085,
  September 2011.

\bibitem{DiSP}
D.~Sundman, S.~Chatterjee, and M.~Skoglund, ``{A greedy pursuit algorithm for
  distributed compressed sensing},'' in \emph{Proc. IEEE Int. Conf. on Acoust.,
  Speech, and Sig. Proc. (ICASSP)}, 2012, pp. 2729--2732.

\bibitem{DBP}
J.~Mota, J.~Xavier, P.~Aguiar, and M.~Puschel, ``{Distributed basis pursuit},''
  \emph{IEEE Trans. Sig. Proc.}, vol.~60, pp. 1942--1956, April 2012.

\bibitem{DIHT}
S.~Patterson, Y.~Eldar, and I.~Keidar, ``{Distributed sparse signal recovery
  for sensor networks},'' in \emph{Proc. IEEE Int. Conf. on Acoust., Speech,
  and Sig. Proc. (ICASSP)}, 2013, pp. 4494--4498.

\bibitem{TA}
R.~Fagin, A.~Lotem, and M.~Naor, ``{Optimal aggregation algorithms for
  middleware},'' in \emph{Symposium on Principles of Database Systems}, 2001,
  pp. 614--656.

\bibitem{AMP}
D.~L. Donoho, A.~Maleki, and A.~Montanari, ``{Message passing algorithms for
  compressed sensing},'' in \emph{Proc. Natl. Acad. Sci.}, vol. 106, Madrid,
  Spain, September 2009, pp. 18\,914--18\,919.

\bibitem{convergence}
A.~Maleki and R.~G. Baraniuk, ``{Least favorable compressed sensing problems
  for the first order methods},'' in \emph{Proc. IEEE Int. Symp. Inf. Theory},
  2011, pp. 134--138.

\bibitem{TUNE}
D.~L. Donoho, A.~Maleki, and A.~Montanari, ``{The Noise-Sensitivity Phase
  Transition in Compressed Sensing},'' \emph{IEEE Trans. Info. Theory},
  vol.~57, pp. 6920--6941, October 2011.

\bibitem{TPUT}
P.~Cao and Z.~Wang, ``{Efficient top-k query calculation in distributed
  networks},'' in \emph{Intl. Symposium on Principles Of Distributed Computing
  (PODC)}, 2004, pp. 206--215.

\bibitem{CAMPDetect}
L.~Anitori, A.~Maleki, M.~Otten, R.~G. Baraniuk, and P.~Hoogeboom, ``{Design
  and Analysis of Compressed Sensing Radar Detectors},'' \emph{IEEE Trans.
  Signal Proc.}, vol.~61, pp. 813--827, February 2013.

\bibitem{Dynamic}
M.~Bayati and A.~Montanari, ``{The Dynamics of Message Passing on Dense Graphs,
  with Applications to Compressed Sensing},'' \emph{IEEE Trans. Info. Theory},
  vol.~57, pp. 764--785, February 2011.

\end{thebibliography}
\end{document}